\documentclass{aipproc}
\usepackage{epsfig}
\usepackage{graphicx}
\layoutstyle{6x9}

\begin{document}
\title{High Resolution Spectroscopy of 4U 1728-34
from a Simultaneous Chandra-RXTE Observation}
\author{A. D'A{\'\i}}{address={Dipartimento di Scienze Fisiche ed Astronomiche, 
Universit\`a di Palermo, via Archirafi n.36, 90123 Palermo, Italy.}}
\author{T. Di Salvo}{address={Dipartimento di Scienze Fisiche ed Astronomiche, 
Universit\`a di Palermo, via Archirafi n.36, 90123 Palermo, Italy.}}
\author{R. Iaria}{address={Dipartimento di Scienze Fisiche ed Astronomiche, 
Universit\`a di Palermo, via Archirafi n.36, 90123 Palermo, Italy.}}
\author{G. Lavagetto}{address={Dipartimento di Scienze Fisiche ed Astronomiche, 
Universit\`a di Palermo, via Archirafi n.36, 90123 Palermo, Italy.}}
\author{N. R. Robba}{address={Dipartimento di Scienze Fisiche ed Astronomiche, 
Universit\`a di Palermo, via Archirafi n.36, 90123 Palermo, Italy.}}
\author{L. Burderi}{address={Osservatorio Astronomico di Roma, Via Frascati 33, 
00040 Monteporzio Catone (Roma), Italy.}}
\author{M. Mendez}{address={SRON National Institute for Space Research, 
Sorbonnelaan 2, 3584 CA Utrecht, the Netherlands}}
\author{M. van der Klis}{address={Astronomical Institute "Anton Pannekoek," 
University of Amsterdam and Center for High-Energy Astrophysics,
Kruislaan 403, NL 1098 SJ Amsterdam, the Netherlands}}
\begin{abstract}
  We  report  on  a simultaneous  Chandra  and  RossiXTE
  observation  of the LMXB  atoll bursting  source 4U  1728-34 
  performed on 2002 March 3--5. We fitted the 1.2--35 keV  continuum 
  spectrum with a blackbody plus a Comptonized component.  
  An overabundance of Si by a factor of $\sim 2$ with respect to Solar
  abundance is required for a satisfactory fit. 
  Large residuals at 6--10 keV can be fitted by a broad (FWHM $\simeq 1.6$
  keV) Gaussian emission line, or, alternatively, by absorption edges 
  associated with Fe I and Fe XXV at $\sim 7.1$ keV and  $\sim 9$ keV, 
  respectively.  In  this interpretation,  we find  no  evidence of broad, 
  or narrow Fe K$\alpha$ line,  between 6 and 7 keV.  
  We tested our alternative modeling of the iron K$\alpha$ region by
  reanalyzing a previous BeppoSAX observation of 4U 1728-34, finding a
  general agreement with our new spectral model.
\end{abstract}
\maketitle
\section{Introduction}
4U 1728-34  is a  well known  prototype of the  class of  the bursting
atoll sources  (\cite{hasinger89}). This was one of  the first sources
to display  kHz QPOs in its power  spectrum and the  first one to
display  burst  oscillations, around  363 Hz,  during some  type I
X-ray bursts  (\cite{strohmayer96}).  Its temporal  behavior has been
recently extensively  studied (\cite{disalvo01}, \cite{vanstraaten02})
using  a large  set of  RXTE  observations, spanning  more than  three
years.  Spectral  studies have been carried  out in the  past with the
use of  EXOSAT (\cite{white86}), RXTE  (\cite{piraino00}) and BeppoSAX
(\cite{disalvo00},   \cite{piraino00})    satellites; although
there was a general agreement about the modelization of the continuum 
emission, it was disputed  the nature of  local features,  especially 
a  broad Gaussian line at  6.2-6.7 keV.    The distance to the source 
is estimated between  4.1 kpc and 5.1 kpc 
(\cite{galloway03}; \cite{disalvo00}).

In this paper we present the results from a simultaneous Chandra-RXTE
observation of 4U 1728-34. We propose an alternative fit of the iron 
K-shell region, using two absorption edges instead of a broad emission
line, whose interpretation would be quite problematic.

\section{Observations and Spectral Analysis}

4U  1728-34 was  observed  by Chandra  on  2002 March  4  for a  total
collecting time  of $\sim 30$ ksec.   We used the  Chandra High Energy
Transmission Grating  Spectrometer (HETGS) to perform  a high resolution
spectroscopic  analysis.   The  data  were collected  in  the  Timed
Exposure Mode, and a sub-array was adopted  (q = 1 and n = 400, see the
Chandra         Proposer's         Observatory        Guide         at
\textrm{http://cxc.harvard.edu/proposer/POG})  in  order to  mitigate
the effects  of photon pile-up for first  order spectra.  Consequently
the frame time  was 1.44 s, the High Energy  Grating (HEG) spectrum is
cut below 1.6 keV and the Medium Energy Grating (MEG) spectrum below
1.2 keV. No systematic error was added to the data.
The RXTE  observation started  on 2002 March  3 03:27:12 and  ended on
2002 March  5 13:00:00.   For the spectral  analysis we used  only the
Proportional Counter Units 2 and 3 data in the Standard2 configuration
(with 16 s time resolution and 129 energy channels).
Seven type-I bursts were revealed in the PCA lightcurve and two bursts
in the Chandra lightcurve.  As our  primary concern is to focus on the
persistent emission of the source we discarded data around each burst,
for  a time length of  160 s.

As concerns the Chandra HETG  data, we considered the four first-order
dispersed spectra, namely the two  HEG spectra and the two MEG spectra
on the opposite sides of  the zeroth order.  We averaged HEG+1 (MEG+1)
and HEG-1  (MEG-1) spectra in  a single spectrum, after  having tested
their reciprocal consistency.  The used energy range is 1.6-10 keV for
the HEG spectrum and 1.2-5 keV for the MEG spectrum.  For all the fits
we took  into account an instrumental  feature at 2.07  keV for bright
sources (described by \cite{miller02}) and fit it with an inverse edge
(with optical depth $\tau \simeq  -0.1$). The HEG and MEG spectra were
binned  in order  to have  at least  300 counts  for each  bin.  This,
however, still ensures a high number of channels (about 1000) and good 
spectral resolution throughout the entire covered energy band.
Concerning  the RXTE/PCA  data the standard selection  criteria for
obtaining  the Good Time  Intervals were  applied.  We  restricted the
spectral analysis to the  temporal intervals during which RXTE operated
simultaneously with  Chandra.  We limited  the energy range  to 3.5-35
keV, and applied a 2\% systematic error for channels below 25 keV and
2.5\% for channels above  25 keV.
Relative normalizations of the three instruments, except for HEG which
was fixed to a reference value of 1, were left as free parameters
in  all the fits performed.  

We tried  a series of models  to simultaneously fit HEG,  MEG and RXTE
spectra.  We found  the best-fit model to consist  of a soft emission,
described  by a  blackbody of  temperature $\simeq$  0.52 keV,  plus a
Comptonized  component (CompTT  in XSPEC,  \cite{titarchuck94}), where
the seed photon  temperature kT$_0$ is $\simeq$ 1.3  keV, the electron
temperature kT$_e$ is $\simeq$ 7.4 keV and, finally, the optical depth
$\tau$  associated to  a  spherical corona  is  $\simeq$ 6.2.   Both
components  are photo-electrically absorbed  by an  equivalent hydrogen
column  $N_H  \simeq 2.3  \times  10^{22}$~cm$^{-2}$.  The  associated
$\chi^2$/d.o.f.  obtained  for  this  fit  is  1206/1034.   We  noted  an
evident absorption edge around 1.84 keV, probably associated to  neutral Si,
in the MEG and HEG spectra. To fit this edge we substituted the component
{\it phabs} in XSPEC with the component {\it vphabs}, which allows us
to vary  the abundances of single  elements with respect  to the solar
abundances.  We found that leaving  the Si abundance free improves the
fits significantly; Si resulted overabundant by a factor $\sim$ 2 with
respect to  the solar abundance ($\chi^2$/d.o.f. value  obtained for this
fit is 1031/1033).

Large residuals at 6--10 keV can be fitted by a very broad Gaussian emission
line ($\sigma \simeq 0.7$ keV, corresponding to a FWHM $\simeq 1.6$ keV), 
whose interpretation is quite problematic,
or, alternatively, by absorption edges associated with Fe I and Fe XXV 
at $7.03$ keV ($\tau$  $\simeq$ 0.11) and  $\sim 9$ keV ($\tau$ $\simeq$
0.16, respectively.  The addition  of these  edges improves  the fit
significantly  compared  to the  simple  model  described above  (giving a
decrease of $\chi^2$/d.o.f. from 1031/1033 to  959/1029).  In  this
interpretation,  we find  no  evidence of broad, or narrow Fe K$\alpha$ line,
between 6 and 7 keV.  To test the  consistency of our
model we reanalyzed a  previous BeppoSAX observation performed between
August 23 and 24, 1998  (see \cite{disalvo00}).  A Gaussian emission line
is  no longer statistically  required if  we introduce  two absorption
edges at energies above 7 keV.   We found a first edge at $\simeq$ 7.4
keV  ($\tau$ $\simeq$  0.08) and  a second  edge at  $\simeq$  8.7 keV
($\tau$  $\simeq$  0.06). We found a general agreement
with  the  Chandra-RXTE  spectrum,  obtaining $\chi^2$/d.o.f.  =  225/179
(instead of $\chi^2$/d.o.f.  = 236/178 obtained for the  model adopted in
\cite{disalvo00}). 
In table 1 and Figure 1 we  present the results of the two fits of the 
BeppoSAX and Chandra-RXTE datasets.  

\section{Conclusions}

We  confirm that  the  best-fit  continuum model  for  this source  is
composed of two components:  a blackbody emission which probably comes
from the  inner edge of an  accretion disk around  the compact object,
and a  Comptonized component coming  from a hot corona  surrounding the
system.    
Although we cannot  definitely exclude that a quite broad  Gaussian line  
($\sigma \simeq 0.7$ keV, FWHM $\simeq 1.6$ keV) is  present in  this
source, we present here a different interpretation of the residuals in
the  iron K-shell  region.  These  are well  fitted by  two absorption
edges at energies  at $\sim 7$ and $\sim 9$ keV,  associated with Fe I
and  Fe  XXV,  respectively  (see e.g. \cite{turner92}).   We  derived,
however, an upper  limit to the flux of a broad  emission line, fixing the
values of  energy and width  to the values found  in \cite{disalvo00};
we find an upper limit to the line flux of $1.7 \times 10^{-11}$ erg 
cm$^{-2}$ sec$^{-1}$ (corresponding to an equivalent width of 28 eV).

We checked that this alternative model is  also in agreement with a previous 
BeppoSAX observation.  We observe a shift in  the energy of the Fe I edge 
from 7.1  keV during the  Chandra-RXTE observation  to 7.4  keV (compatible
with  K$\alpha$ edges of  moderately ionized  iron, Fe  IX to  Fe XVI)
during the BeppoSAX  observation, while the energy of  the Fe XXV edge
is consistent  with being the  same in both observations.   Finally we
found,  both   in  the  Chandra-RXTE   and  BeppoSAX  spectra,   a  Si
overabundance  by a  factor $\sim  2-2.5$  with respect  to the  solar
abundance and  the values  obtained from the  fit are  consistently in
agreement.

\begin{table}
\centering
\caption{ \footnotesize  \linespread{1}
  Best fit parameters for the two datasets of 4U 1728-34,
  obtained from a BeppoSAX observation (0.12-60 keV energy band)
  and a joint CHANDRA-RXTE observation (1.2-35 keV energy band).
  The continuum emission consists of a thermal blackbody
  (bbody) and a Comptonized component modeled by compTT.
  $kT_{\rm BB}$  and
  N$_{\rm  BB}$  are,  respectively,  the  blackbody  temperature  and
  normalization in  units of $L_{39}/D_{10}^2$, where  $L_{39}$ is the
  luminosity in units of $10^{39}$ ergs/s and $D_{10}$ is the distance
  in  units  of  10  kpc.   $kT_0$, $kT_{\rm e}$  and  $\tau$  indicate  the
  seed-photon  temperature, the electron  temperature and  the optical
  depth  of the Comptonizing  cloud around  the neutron  star. N$_{\rm
  Comptt}$  is  the  normalization  of  the Comptt  model  in  XSPEC
  v.11.2.0 units. Unabsorbed luminosities of the bbody component and of the 
  CompTT component are calculated assuming a distance to the source of 5.1 kpc 
  (\cite{disalvo00}).
  For the component Edge, E$_{edge}$ denotes the energy of the edge and 
  $\tau$ the optical depth.
  Uncertainties are  at 90\% confidence level for  a single parameter.}
\label{tab_sax2}
\begin{tabular}{l l l |l}
  \hline
  \hline
  &                  & BeppoSAX       & Chandra - RXTE  \\
  \hline
  Component & Parameter (Units) & \multicolumn{2}{c}{Values} \\
  \hline
  vpha  &  $N_{\rm  H}$    $(10^{22}$  cm$^{-2})$   & $2.37^{+0.14}_{-0.11}$ &   $2.61^{+0.06}_{-0.07}$     \\
  vpha  & Si (Solar units)      & $2.1_{-0.5}^{+0.5}$    &   $2.02_{-0.13}^{+0.13}$      \\


  edge       & E$_{edge}$     (keV)      & $7.41^{+0.15}_{-0.14}$     & $7.03^{+0.08}_{-0.06}$  \\
  edge       &  $\tau$           $(10^{-2})$   & $8^{+2}_{-2}$  & $11^{+3}_{-4}$    \\

  edge       & E$_{edge}$     (keV)      & $8.73^{+0.26}_{-0.24}$       & $9.0^{+0.3}_{-0.4}$ \\
  edge       &     $\tau$        $(10^{-2})$   & $6^{+3}_{-3}$   & $16^{+4}_{-4}$  \\

  bbody      & kT        (keV)      & $0.573_{-0.024}^{+0.002}$      & $0.516_{-0.014}^{+0.002}$  \\
  bbody      & N$_{\rm  BB}$         $(10^{-3})$     & $21.5_{-1.3}^{+1.6}$  & $10.3_{-0.4}^{+0.4}$ \\
  bbody & Luminosity   $(10^{36}$ erg cm$^{-2}$ sec$^{-1})$        & $5.7$ & $0.87$ \\

  CompTT     & $kT_0$ (keV)                 & $1.53^{+0.06}_{-0.07}$    & $1.33^{+0.05}_{-0.05}$  \\
  CompTT     & $kT_{\rm e}$ (keV)                 & $6.4_{-0.4}^{+1.3}$       & $7.4_{-0.4}^{+0.5}$  \\
  CompTT     & $\tau$                       & $4.8^{+0.8}_{-1.0}$       & $6.2^{+0.4}_{-0.6}$   \\
  CompTT     & N$_{\rm CompTT}$ $(10^{-2})$ & $6.8^{+1.3}_{-1.4}$ & $4.4^{+0.3}_{-0.6}$    \\
  CompTT & Luminosity  $(10^{36}$ erg cm$^{-2}$ sec$^{-1})$ &   10.2 & 2.6     \\
         & $\chi^2$/dof              & 225/179              &    959/1029      \\
  \hline
\end{tabular}
\end{table}
\begin{figure}[h!]
\centering
\begin{tabular}{l l}
\epsfig{figure=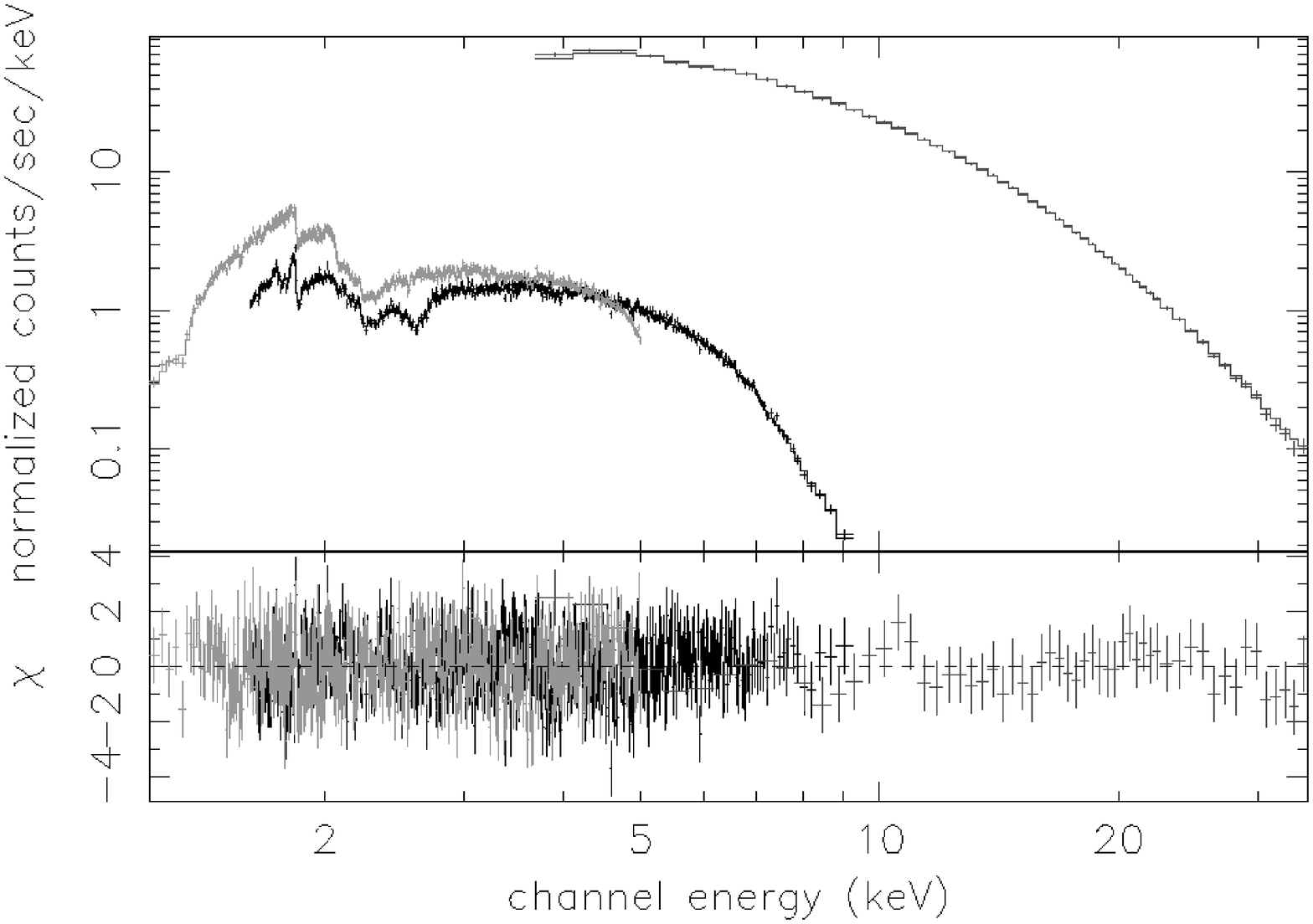, height=3.9cm} & \epsfig{figure=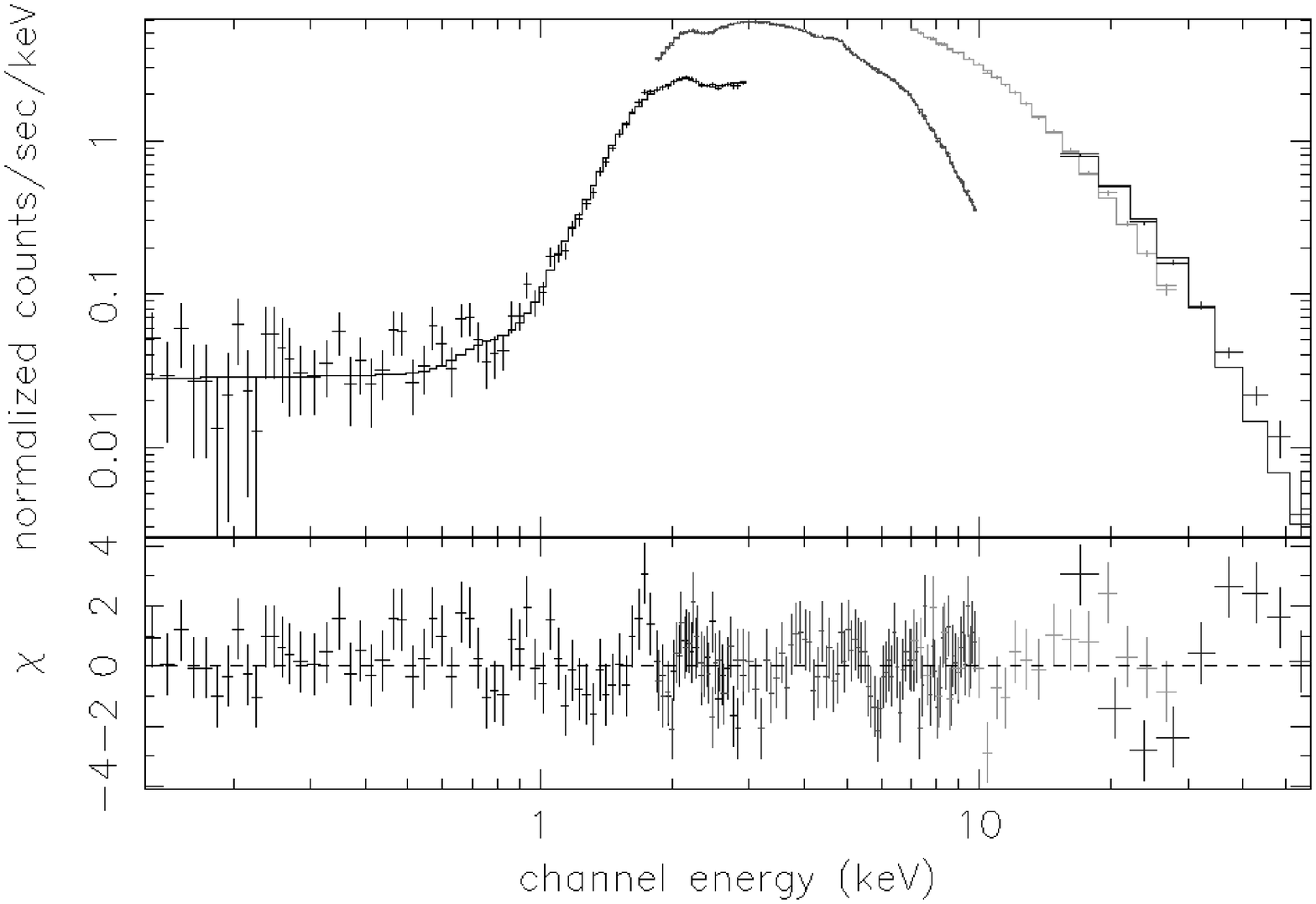, height=3.9cm}\\
\end{tabular}
\caption{  \footnotesize \linespread{1}  Spectra of  4U  1728-34 shown
  together with the best-fit model (see Table 1). Left panel: 1.2--35.0
  keV spectrum obtained from the simultaneous Chandra and RXTE dataset.
  Right panel: 0.12--60  keV spectrum  obtained from a BeppoSAX observation.  
  In smaller  panels: residuals in unit of $\sigma$
  with respect to the best-fit model.}
\label{spettro_chandrasax}
\end{figure}


\begin{thebibliography}{00}
\bibitem{disalvo00} Di Salvo T., Iaria R., Burderi L., Robba N.R., 2000, ApJ, 542, 1034
\bibitem{disalvo01} Di Salvo T., Mendez M., van der Klis M., Ford E., Robba N.R., 2001, ApJ, 546, 1107
\bibitem{galloway03} Galloway D.K., Psaltis D., Chakrabarty D., Muno M.P., 2003, ApJ, 590, 999
\bibitem{hasinger89} Hasinger, G., \& van der Klis, M., 1989, A\&A, 225, 79
\bibitem{miller02} Miller J.M., Fabian A.C., Wijnands R., et al., 2002, ApJ, 578, 450
\bibitem{piraino00} Piraino S., Santangelo A., Kaaret P., 2000, A\&A, 360, L35
\bibitem{strohmayer96} Strohmayer T.E., Zhang W., Swank J.H., et al., 1996, ApJ, 469, L9
\bibitem{titarchuck94} Titarchuck L., 1994, ApJ, 434, 570
\bibitem{turner92} Turner T.J., Done C., Mushotzky R., Madejski G., Kunieda H., 1992, ApJ, 391, 102
\bibitem{vanstraaten02} van Straaten S., van der Klis M., Di Salvo T., Belloni T., 2002, ApJ, 568, 912
\bibitem{white86} White N.E., Peacock A., Hasinger G., et al., 1986, MNRAS, 218, 129
\end{thebibliography}
\end{document}